# An Open, Multi-Platform Software Architecture for Online Education in the Metaverse

Santiago Lombeyda
Center for Data-Driven Discovery,
California Institute of Technology,
Pasadena, California, USA
santiago@caltech.edu

S. George Djorgovski
Center for Data-Driven Discovery
and Astronomy Dept., California
Institute of Technology, Pasadena,
California, USA
djorgovski@caltech.edu

An Tran
California Institute of Technology,
Pasadena, California, USA
antran@caltech.edu

Joy Liu
California Institute of Technology,
Pasadena, California, USA
jliu7@caltech.edu

Alison Noyes
California Institute of Technology,
Pasadena, California, USA
anoyes@caltech.edu

Sasha Fomina
California Institute of Technology,
Pasadena, California, USA
afomina@caltech.edu

## ABSTRACT
Use of online platforms for education is a vibrant and growing arena, incorporating a variety of software platforms and technologies, including various modalities of extended reality. We present our Enhanced Reality Teaching Concierge, an open networking hub architected to enable efficient and easy connectivity between a wide variety of services or applications to a wide variety of clients, designed to showcase 3D for academic purposes across web technologies, virtual reality, and even virtual worlds. The agnostic nature of the system, paired with efficient architecture, and simple and open protocols furnishes an ecosystem that can easily be tailored to maximize the innate characteristics of each 3D display environment while sharing common data and control systems with the ultimate goal of a seamless, expandable, nimble education metaverse.

## 1 INTRODUCTION

3D content can be key in understanding key physical and structural characteristics of a countless number of models across academia. There is a growing number of tools for education that include interactive 3D models. However, to be successful each of these tools need to be carefully crafted to the platform to be utilized to foment smooth and intuitive interaction between the end user and the information being presented. Tools developed to work over traditional web browsers, tools working inside semi-immersive Virtual Worlds (VWs), and tools meant to work in Virtual Reality (VR) or Mixed Reality (MR) need to carefully consider the particular viewpoint(s) of the user, the interaction modalities, the resolution of the display, the ability to clearly distinguish 3D objects or read 2D text or graphs, among many other design factors. Furthermore, as most of these technologies are capable of building in meaningful synchronous and asynchronous collaborative paradigms, they need to factor in the complexities and nuances of each particular virtual space when it comes to how other virtual participants interact with the same information or view each other within the created spaces. While there is a propensity to attempt to design an all-encompassing optimal single space, with each problem, each set of 3D models, each goal may require custom solutions. Thus, we must strive not to create monolithic solutions, but rather to create the dynamic modules that will allow us to scaffold and build the needed tools. In the process, with a mindset of nimbleness, we can start creating solutions that can in fact be easily adapted to work across different types of platforms, while remaining efficient and true to the strengths of each environment.

One application arena of a particular interest and a great societal value is education. We present our Enhanced Reality Teaching (ERT) Concierge as a server-side service that is both agnostic to clients and applications, allowing different flavors of both to work concurrently and efficiently.

### 1.1 Background of Education in Collaborative Spaces

Like many other industries, education is actively being transformed by the Internet. Online education offers the possibilities of a personalized, affordable, and life-long education, a rich variety of material, along with the conveniences of the places and times of the learner's choosing. However, while the advent of online video lectures has solved the challenge of a scalable educational content delivery, some key challenges remain, e.g., the social components of learning or the effective hands-on learning labs. Development of the various forms of virtual (VR) or extended (XR) reality platforms offer potential solutions to both challenges.

There has been a considerable exploration along these lines, initially mainly using Virtual Worlds (VWs), shared multi-user virtual environments that can be accessed through the traditional 2D devices, but offer a reasonable approximation of a subjective 3D immersive experience. There is substantial literature about the use and effectiveness of VWs as an educational platform; see, e.g., (Mikropoulos and Natsis, 2011, van der Land et al., 2011, Djorgovski et al., 2013, Merchant et al., 2014, Freina and Ott, 2015, Bower et al., 2017), among many others. A consensus seems to be that VWs can lead to better learning outcomes, as the students find the lectures and demonstrations more engaging and more memorable than those delivered through the standard video or conferencing platforms. At the same time VWs provide a channel for the human interaction component of the learning process (both student-to-student and student-to-instructor) that is clearly superior to other online mechanisms.

XR is particularly well suited for the virtual hands-on labs, as it gives a more visceral sense than the standard point/click/swish interactions for flat screen devices. One can ideate and build models and experiments in VR that are simply not possible in real life (e.g., construct a full-scale space station; explore a living system from the inside out; create a new galaxy and experience its evolution over cosmic timescales); and there are no physical dangers (e.g., an exploding apparatus). It is also a natural collaborative environment, where multiple users who can be physically in different locations can interact with each other and with data in a shared virtual space. It is a natural environment for the generations of "digital native" students.

As the technologies evolve, and we approach a future 3D Web, with effective and ubiquitous XR interfaces, often referred to as the Metaverse, online education will also evolve; see, e.g., (Hirsh-Pasek et al., 2022, Suzuki et al., 2022, Tlili et al., 2022, Akour et al., 2022), etc. Given the rapid technology evolution in this arena, it may be unlikely that a single platform would dominate. Even today, students and educators use a variety of the Web-based tools and platforms.

With that in mind, we have been developing an experimental, multi-platform educational software environment that would enable customizable implementations, including immersive 3D spaces as display and interaction venues. We present a server-side service that is both agnostic to clients and applications, allowing different flavors of both to work concurrently and efficiently.

## 2 THE ERT CONCIERGE

We have been exploring the potential of XR as an educational platform through our Enhanced Reality Teaching (ERT) project (Lombeyda et al., 2019). The ERT Concierge is a simple and efficient server-side networking hub, which treats anyone connecting to it as an equal participant (ERTConcierge-GitHub, 2022) as depicted in

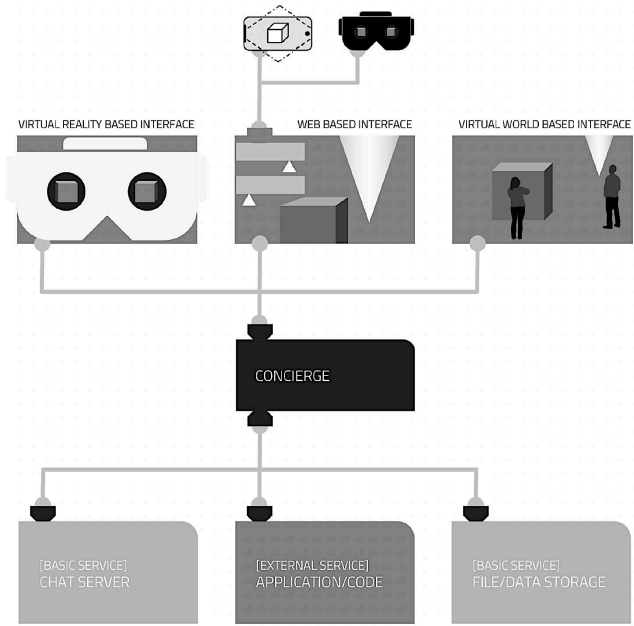

**Figure 1: A schematic configuration of the ERT Concierge system.**

Figure 1. However, as application services initiate communication with the ERT Concierge they can register for special allowances as to who can communicate with, who they reply to, and the type of information being conveyed. This is very similar to a modern chat-like service, like Discord, where users sign in to special servers, where they are given specific group permissions as to whom they can talk to or not; and equally bots can connect to these networks and offer all sorts of services, ranging from music streaming bots to virtual assistant bots, to bots offering fully responsive role-playing game interactions.

ERT Concierge is a central server, written in the Rust programming language and built upon Actix, one of the fastest web frameworks as of 2020 (Actix, 2022). The protocol consists of a native WebSocket relay and an HTTP file server.

The socket relay uses JavaScript Object Notation (JSON), an industry-standard and human-readable data serialization format, as the primary method of communication between connected clients and the server. The client can send a variety of commands to the central server, allowing for retrieval of data or communication with other connected clients. The networking is done over web sockets, which is well suited for extensibility and maintainability since libraries and support for it are ubiquitous in every programming language and environment. These rely on a supplemental HTTP file server that allows for transfers of complex and large files, like meshes or computational datasets, over the network. It supports multipart uploading and streaming downloads, which is compatible with a variety of HTTP implementations and browsers.

### 2.1 ERT Concierge + Web + VR

As an example, a planetary system simulator that incorporates the appropriate gravitational interactions was written using Python.

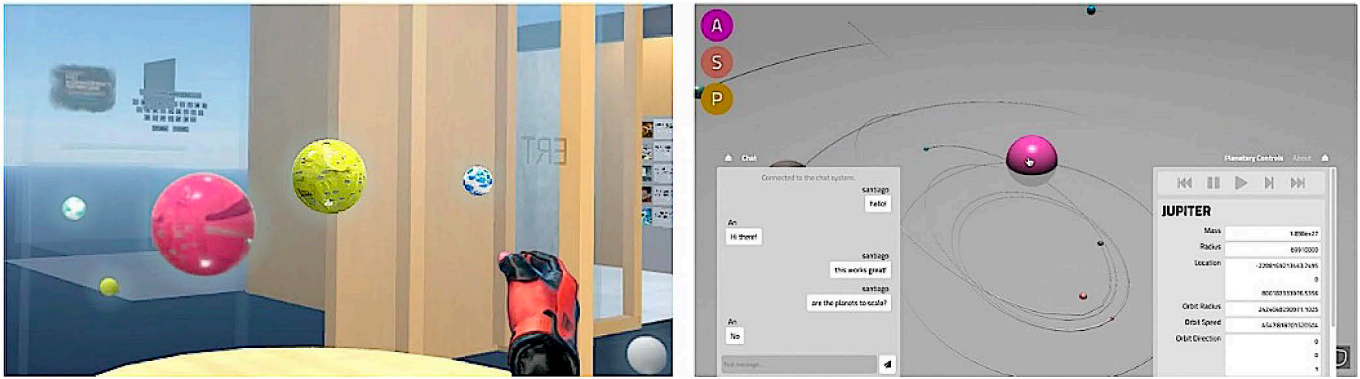

**Figure 2: Planetary simulation running in VR on the left, and on a browser, with distinct UI methodologies, but same concurrent data**

Results of the planetary simulation were then transmitted to both a Babylon.js (BabylonJS, 2022) web client, as well as Unity (Unity, 2022) based collaborative classroom (Fig. 2).

Similarly, a Rust physics simulation of freely moving colliding boxes was written as a standalone physics engine. The standalone physics engine displayed its state using Simple DirectMedia Layer (SDL2) (SDL2, 2022) on the native machine. An adapter for the central server was written that sent state information of the boxes to the server. The state contains only positional and identification information for each box. Clients must then request the exact specifications of the boxes' dimensions, colors, and IDs to properly draw them. In the end, the simulation streamed data to the client in real-time (~50 state updates per second), to which the BabylonJS client reacts to and appropriately draws the 3D model of the simulated 2D world.

## 2.2 An example of the ERT Concierge using Second Life as a 3D display and interaction space

As an example of a use case, we employ the ERT Concierge to connect an external physics-based simulation to a 3D environment of the VW Second Life (SL) (Second Life, 2022). Corrade (Corrade, 2022), a free and open source bot that is placed in the VW, and stays connected to the SL server grid while running, became our best available choice to connect SL to the ERT Concierge while giving us access to a fully supported implementation of Web Sockets. Corrade acts like a bridge, allowing users to send commands through socket communication to the bot, which relays the information to SL. Figure 3 shows Corrade supporting an external toy aquarium React (React, 2022) app, which ultimately allows the external interaction with SL from an external website. The React app allows users to create and delete sea creatures/plants with the click of a button.

While working with Corrade to support the two-way communication posed a great challenge. The key to enabling usability with SL was the use of Universal Unique Identifiers (UUID), a unique string that can be used to identify 'in-world' objects. The UUID, as created by the bot, would enable tracking of in-world objects by the external application in order to affect their properties. Initial tests included moving these objects to different positions, either by typing in specific coordinates or by incrementing the object's position in the x, y, or z directions. This aquarium React app acted as a demonstration of the abilities of Corrade and probed the extent of which socket communication could operate in SL.

A more complex example included the creation of a physics server utilizing the popular online physics engine PyBullet (PyBullet, 2022), and clients on a web browser utilizing Babylon and in SL to furnish a web-based UI for a physics-based game running in the virtual world. The physics engine was built to dictate the movement of objects across all front ends. PyBullet was the most suitable for easy wrapping and setting up connection with the ERT Concierge for our purposes.

## 3 CONCLUSIONS

We have demonstrated that the architecture behind the ERT Concierge may be the key behind an easier path to create platform-agnostic collaborative educational experiences with richer and more diverse functionalities. Such a system can empower educators and education technologists to easily wrap applications and code bases as services that can easily connect via web sockets to the ERT Concierge, complimented with tested 3D clients on the Web, VR, and VWs of their choice. Furthermore, this software ecosystem can easily be expanded to create single experiences that expand multiple platforms even for one single user. This flexible approach can facilitate educational and other uses of the emerging Metaverse, as the variety of its components and software platforms continue to increase in both the number and the complexity.

Our hope is that as we further improve our ERT Concierge, we will amalgamate a larger collection of services which can easily be added to a collaborative space, as well as curate and offer UI templates for the different end display client environments that make it simple to create experiences that work across platforms, but are capable of exploiting any environments' innate traits.

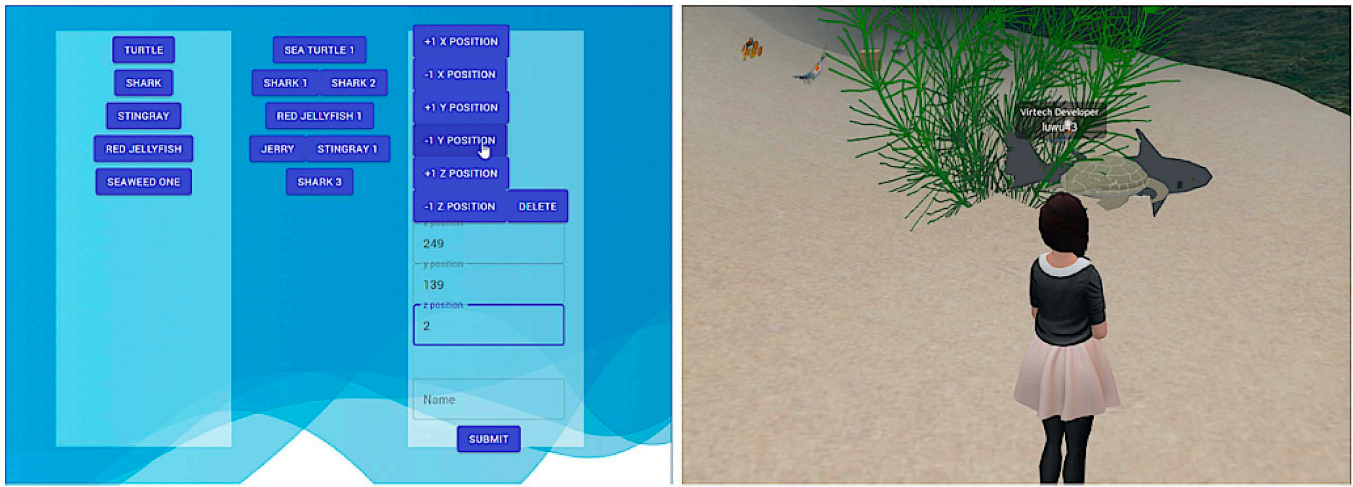

Figure 3: External browser (left) as a UI interacting with the rendered 3D scene in Second Life.

## ACKNOWLEDGMENTS
ERT ecosystem expands on the ERT VR Classroom (Lombeyda et al., 2019). Work for ERT and ERT Concierge was partially funded by Caltech Provost's Innovation in Education Fund, and Caltech's Summer Research Fellowship Program. Hardware donations from HTC Vive and NVIDIA.

## REFERENCES

Tassos A. Mikropoulos, Antonis Natsis. 2011. Educational virtual environments: A ten-year review of empirical research (1999–2009). Computers & Education, 56(3), 769-780. https://doi.org/10.1016/j.compedu.2010.10.020

Sarah van der Land, Alexander P. Schouten, Bart van den Hooff, Frans Feldberg. 2011. Modelling the Metaverse: A Theoretical Model of Effective Team Collaboration in 3D Virtual Environments. Journal of Virtual Worlds Research, 4(3), 1-16. https://doi.org/10.4101/jvwr.v4i3.6126

S. George Djorgovski, Piet Hut, Rob Knop, Giuseppe Longo, Steve McMillan, Enrico Vesperini, Ciro Donalek, Matthew Graham, Ashish Mahabal, Franz Sauer, Charles White, Crista Lopes. 2013. In SLACTIONS 2012 International Research Conference on Virtual Worlds, eds. L. Morgado, Y. Sivan, A.M. Maia, *et al.*, pp. 49-58. Vila Real, PT: UTAD. https://repositorio.utad.pt/handle/10348/2305 or https://arxiv.org/abs/1301.6808

Zahira Merchant, Ernest T. Goetz, Lauren Cifuentes, Wendy Keeney-Kennicutt, Trina J. Davis. 2014. Computers & Education, 70, 29-40. https://doi.org/10.1016/j.compedu.2013.07.033

Laura Freina, Michela Ott. 2015. A Literature Review on Immersive Virtual Reality in Education: State Of The Art and Perspectives. In eLearning and Software for Education (eLSE), Vol. 1, 133-141. https://doi.org/10.12753/2066-026X-15-020

Matt Bower, Mark J. W. Lee, Barney Dalgarno. 2017. Collaborative learning across physical and virtual worlds: Factors supporting and constraining learners in a blended reality environment. British Journal of Educational Technology, 48(2), 407-430. https://doi.org/10.1111/bjet.12435

Kathy Hirsh-Pasek, Jennifer M. Zosh, Helen S. Hadani, Roberta M. Golinkoff, Kevin Clark, Chip Donohue, Ellen Wartella. 2022. A whole new world: Education meets the Metaverse. Brookings Institution report. https://www.brookings.edu/research/a-whole-new-world-education-meets-the-metaverse/

Sin-nosuke Suzuki, Hideyuki Kanematsu, Dana M. Barry, Nobuyuki Ogawa, Kuniaki Ya- jima, Katsuko T. Nakahira, Tatsuya Shirai, Masashi Kawaguchi, Toshiro Kobayashi, Michiko Yoshitake. 2022. Virtual Experiments in Metaverse and their Applications to Collaborative Projects: The framework and its significance. Procedia Computer Science, 176, 2125-2132. https://doi.org/10.1016/j.procs.2020.09.249

Ahmed Tlili, Ronghuai Huang, Boulus Shehata, Dejian Liu, Jialu Zhao, Ahmed Hosny, Saleh Metwally, Huanhuan Wang, Mouna Denden, Aras Bozkurt, Lik-Hang Lee, Dogus Beyoglu, Fahriye Altinay, Ramesh C. Sharma, Zehra Altinay, Zhisheng Li, Jiahao Liu, Faizan Ahmad, Ying Hu, Soheil Salha, Mourad Abed, Daniel Burgos. 2022. Is Metaverse in education a blessing or a curse: a combined content and bibliometric analysis. Smart Learning Environments, 9:24. https://doi.org/10.1186/s40561-022-00205-x

Iman A. Akour, Rana Saeed Al-Maroof, Raghad Alfaisail, Said A. Salloum. 2022. A conceptual framework for determining metaverse adoption in higher institutions of gulf area: An empirical study using hybrid SEM-ANN approach. Computers and Education: Artificial Intelligence, 3: 100052. https://doi.org/10.1016/j.caeai.2022.100052

Santiago Lombeyda, Lucy Chen, Netra Ravishankar, S. George Djorgovski. 2019. ERT: An Enhanced Reality Teaching Space. In EDUCAUSE ELI Annual Meeting. https://events.educause.edu/eli/annual-meeting/2019/agenda/a-highereducation-virtual-reality-classroom-an-enhanced-reality-for-teaching 2019. Accessed 30 Sep. 2022.

ERTConcierge-GitHub github.com/Avarel/ert-concierge. Accessed 30 Sep. 2022.

Actix www.techempower.com/benchmarks/#section=data-r19. Acc. 30 Sep. 2022.

BabylonJS www.babylonjs.com. Accessed 30 Sep. 2022.

Unity unity.com/. Accessed 30 Sep. 2022.

SDL2 www.libsdl.org/. Accessed 30 Sep. 2022.

Second Life secondlife.com/. Accessed 30 Sep. 2022.

Corrade corrade.grimore.org/. Accessed 30 Sep. 2022.

React reactjs.org/. Accessed 30 Sep. 2022.

PyBullet pybullet.org/. Accessed 30 Sep. 2022.